\documentclass{aa}  

\usepackage{graphicx}
\usepackage{longtable}
\usepackage{lscape}
\usepackage{natbib}
\usepackage{lscape}
\usepackage[varg]{txfonts}
\begin{document}
%
   \title{2MASS\,J154043.42$-$510135.7: a new addition to the 5 pc population}

   \subtitle{ }

   \author{A. P\'erez Garrido \inst{1}\thanks{Based on observations collected at the 
   European Organisation for Astronomical Research in the Southern Hemisphere, Chile.}
          \and
          N.\ Lodieu \inst{2,3}
          \and
          V.\ J.\ S.\ B\'ejar \inst{2,3}
          \and 
          M.\ T.\ Ruiz \inst{4}
          \and
          B.\ Gauza \inst{2,3}
          \and
          R.\ Rebolo \inst{2,3,5}
          \and
          M.\ R.\ Zapatero Osorio \inst{6}
          }

   \institute{Dpto. F\'\i sica Aplicada, Universidad Polit\'ecnica de Cartagena,  Campus Muralla del Mar, Cartagena, Murcia E-30202, Spain \\
         \email{antonio.perez@upct.es}
         \and
         Instituto de Astrof\'isica de Canarias (IAC), Calle V\'ia L\'actea s/n, E-38200 La Laguna, Tenerife, Spain
         \and
         Departamento de Astrof\'isica, Universidad de La Laguna (ULL), E-38205 La Laguna, Tenerife, Spain
         \and
         Departamento de Astronom\'ia, Universidad de Chile, Casilla 36-D, Santiago, Chile
         \and 
         Consejo Superior de Investigaciones Cient\'ificas, CSIC, Spain
         \and
         Centro de Astrobiolog\'ia (CSIC-INTA), Ctra. Ajalvir km 4, 28850, Torrej\'on de Ardoz, Madrid, Spain
         }

   \date{\today{}; \today{}}

 
  \abstract
   {}
   {The aim of the project is to find the stars nearest to the Sun and to contribute
to the completion of the stellar and substellar census of the solar neighbourhood.}
   {We identified a new late-M dwarf within 5 pc, looking for high proper motion sources 
    in the 2MASS--WISE cross-match. We collected astrometric and photometric
   data available from public large-scale surveys. We complemented this
   information with low-resolution (R$\sim$500) optical (600--1000 nm) and 
   near-infrared (900--2500 nm) spectroscopy with instrumentation on the
   European Southern Observatory New Technology Telescope to 
   confirm the nature of our candidate. We also present a high-quality medium-resolution
   VLT/X-shooter spectrum covering the 400 to 2500 nm wavelength range.}
   {We classify this new neighbour as an M7.0$\pm$0.5 dwarf using spectral templates
   from the Sloan Digital Sky Survey and spectral indices. Lithium absorption 
   at 670.8 nm is not detected in the X-shooter spectrum, indicating that the M7 dwarf is 
   older than 600 Myr and more massive than 0.06 M$_{\odot}$. We also derive a trigonometric 
   distance of 4.4$^{+0.5}_{-0.4}$\,pc, in agreement with the spectroscopic distance 
   estimate, making 2MASS\,J154043.42$-$510135.7 (2M1540) the nearest M7 dwarf to the Sun. 
   This trigonometric distance is somewhat closer than the $\sim$6 pc distance 
   reported by the ALLWISE team, who independently identified this object recently.
   This discovery represents an increase of 25\% in the number of M7--M8 dwarfs already 
   known at distances closer than 8\,pc from our Sun. We derive a density of 
   $\rho$\,=\,1.9$\pm$0.9$\times$10$^{-3}$\,pc$^{-3}$ for M7 dwarfs in the 8 pc volume, 
   a value similar to those quoted in the literature.}
   {This new ultracool dwarf is among the 50 nearest systems to the Sun, demonstrating
   that our current knowledge of the stellar census within the 5 pc sample remains 
   incomplete. 2M1540 represents a unique opportunity to search for extrasolar
   planets around ultracool dwarfs due to its proximity and brightness.}
  

   \maketitle
%

%
%
\section{Introduction}
\label{close_dwarf:intro}

The quest for the nearest and coolest neighbours has always been a holy grail because
these objects offer rare opportunities to study stellar properties in great detail 
(e.g.\ atmospheric composition, age, mass, binarity, etc.) 
and they represent unique targets for extrasolar planet searches. The brightest
stars such as Sirius, Alpha Centauri and Procyon have been recorded in 195\,BC, 650\,BC,
and 275\,BC, respectively. Moreover, most of the systems within 5 pc of the Sun have 
been discovered over several centuries. Only a few additions to the list of nearest 
systems have been announced.
Since 2000, sixteen new objects have been added to the list of nearest systems. 
The Wide-field Infrared Survey Explorer \citep[WISE;][]{wright10}
has contributed to some of these discoveries, particularly to those with LTY spectral types 
\citep{kirkpatrick11,cushing11,kirkpatrick12,marsh13,dupuy13b,bihain13a,luhman13a,mamajek13a,marsh13}.
The other discoveries within 5 pc from the Sun that do not originate from WISE are 
the Epsilon Indi Ba/Bb binary at 3.626 pc made of a T1$+$T6 \citep{scholz03,mjm04,king10b}, 
the T6 substellar companion to SCR\,1845$-$6357 \citep{biller06,kasper07a} at 3.84 pc
\citep{deacon05a}, the T9 dwarf UGPS\,0722$-$05 at 4.12 pc \citep{lucas10,leggett12a}, 
the T6 dwarf DENIS\,J081730.0$-$615520 at 4.93 pc \citep{artigau10}.
Two stellar additions to the 5 pc census are SCR\,1845$-$6357 \citep[M8.5;][]{henry06} and
Teegarden's star at 3.84 pc \citep[M6.5;][]{teegarden03}.

With the advent of large-scale surveys, in particular of the Sloan Sky Digital Survey 
\citep[SDSS;][]{york00}, the properties of large samples of M dwarfs were characterized 
\citep{hawley02,west04,schmidt10b}. \citet{bochanski10} reported new measurements of the 
luminosity and mass functions of field low-mass dwarfs derived from SDSS data release 6 (DR6)
over an area of 8400 square degrees. \citet{west11} presented a spectroscopic catalogue 
of 70,841 M dwarfs from the SDSS DR7, including the identification of eight new late-type 
M dwarfs possibly within 25\,pc. However, the density of the latest M dwarfs still suffers 
from many uncertainties, with the first estimate of their density from \citet{kirkpatrick94},
see also the compilation by \citet{caballero08e}. More recent studies, conducted in the solar 
neighbourhood by \citet{cruz07}, yielded a density of 1.9--2.2$\times$10$^{-3}\,pc^{-3}$ 
for M7--M8 dwarfs.

The main objective of our investigation is to uncover overlooked stars and brown dwarfs 
in the solar neighbourhood, by searching for high proper motion sources by cross-correlating 
public databases. In this paper, we report on a fast-moving
late-type M dwarf, 2MASS\,J154043.42$-$510135.7 (hereafter 2M1540), identified 
in a cross-correlation between the Two Micron All-Sky Survey 
\citep[2MASS;][]{cutri03,skrutskie06} and WISE \citep{wright10} public all-sky surveys.
In the process of writing this paper, we learned of the independent discovery of this object  
by Kirkpatrick et al.\ (2014), who reported it as an M6 dwarf located at $\sim$6pc. 
Here we present new astrometry, photometry, and optical and near-infrared spectroscopy, providing 
a slightly different trigonometric distance and spectral type determination.
2M1540 is among the 50 closest systems, the third late-M discovered during the past decade,
and the closest M7 dwarf to the Sun.
In Section \ref{close_dwarf:selection} we describe our selection procedure that
led to the identification of this nearby M dwarf.
In Section \ref{close_dwarf:spectro_followup} we present our optical and near-infrared
spectroscopic follow-up and assign a spectral type to 2M1540\@.
In Section \ref{close_dwarf:distance} we estimate the parallax of 2M1540
using all epochs available in public databases and derive a spectroscopic distance.
In Section \ref{close_dwarf:conclusions} we place our work in context and present
some future work to look for extrasolar planets.

%
%
%
\begin{figure*}
  \includegraphics[width=\linewidth, angle=0]{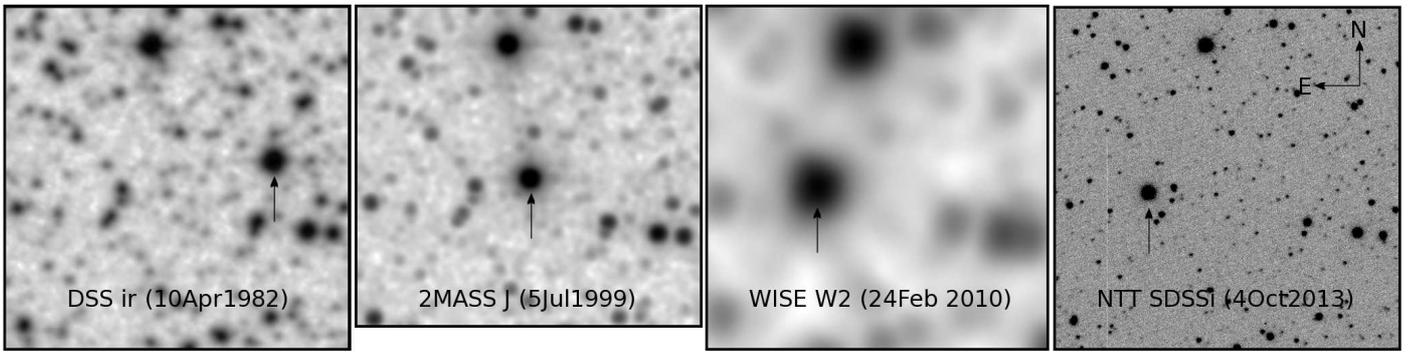}
  \centering
   \caption{Images of the new M dwarf: from left to right are shown
   the optical images from the Digital Sky Survey (DSS\,IR; 10 April 1982),
   the near-infrared $J$-band from 2MASS (5 July 1999), and the mid-infrared
   4.5 micron image from WISE (24 February 2010), and our NTT EFOSC2 image
   (4 October 2013) to illustrate the motion and colours
   of 2M1540\@. Each image is $\sim$2 arcmin aside with North up and East left.
   }
   \label{fig_close_dwarf:images}
\end{figure*}
%

%
%
\section{Selection method}
\label{close_dwarf:selection}

We cross-correlated the 2MASS Point Source Catalogue \citep{cutri03} with the All Sky WISE 
Catalogue \citep{wright10} with the aim of finding red/brown dwarfs in the solar neighbourhood. 
In particular, we focused on objects brighter than $J$\,=\,12 mag. In our search, 
we looked for pairs of uncorrelated objects in 2MASS and WISE that are separated by distances 
larger than two arcsec and less than 60 arcsec between both catalogues. We considered ``uncorrelated'' 
objects those that are in one catalogue with no counterpart in the other database within our
minimum radius of two arcsec. These pairs represent high proper motion candidates with motions
in the range $\sim$0.15--4.6 arcsec/yr, given the temporal baseline between both surveys ($\sim$11--14\,yr). 
A given pair was discarded from our list of candidates if there existed a 
counterpart in the USNO-B1 catalogue \citep{monet03}. Furthermore, we excluded from our list
point sources with 2MASS $J-K_{s}$ colours between 0.4 and 0.8 mag, negative WISE colours
($w1-w2$\,$\leq$\,0 mag), and less than four WISE detections ({\tt{w1nm}} and {\tt{w2nm}} 
criteria less than 4) to focus our search on low-mass stars and/or brown dwarfs. 
Additional contaminants were removed by visual inspection of the multi-epoch images 
(Fig.\ \ref{fig_close_dwarf:images}). 
In our final list, we identified a number of already known nearby M and L dwarfs, 
e.g.\ WISE J104915.57$-$531906.1 at 2 pc \citep[or Luhman16AB;][]{luhman13a,mamajek13a} 
and the late-M dwarf 2M1540, which is the subject of this paper.

%
%
%
\begin{table}
 \centering
 \caption[]{Coordinates, optical, and infrared photometry from public databases and our own measurements, 
 along with our parallax and proper motion measurements and physical properties derived for 2M1540. 
 }
 \begin{tabular}{c c}
 \hline
 \hline
Type  & Value \cr
 \hline
 RA$_{\rm 2MASS}$ (J2000)      & 15:40:43.42  \cr
dec$_{\rm 2MASS}$ (J2000)       & $-$51:01:35.7 \cr
RA$_{\rm WISE}$ (J2000)           & 15:40:45.65 \cr
dec$_{\rm WISE}$ (J2000)          & $-$51:01:39.3 \cr
$\mu_{\alpha}\cos{\delta}$ & $+$1.954$\pm$0.002 arcsec/yr \cr
$\mu_\delta$ & $-$0.330$\pm$0.003 arcsec/yr \cr
$\pi$ &   0.228$\pm$0.024 arcsec \cr
$V$ (EFOSC2)  & 15.26$\pm$0.06 mag \cr
$R_{\rm EFOSC2}$  & 13.23$\pm$0.04 mag \cr
SDSS$r$ (EFOSC2) & 13.88$\pm$0.07 mag \cr
SDSS$i$ (EFOSC2) & 11.60$\pm$0.04 mag \cr
SDSS$z$ (EFOSC2) & 10.76$\pm$0.05 mag \cr 
$J_{\rm 2MASS}$ & 8.961$\pm$0.025 mag \cr 
$H_{\rm 2MASS}$ & 8.299$\pm$0.038 mag \cr 
$Ks_{\rm 2MASS}$ & 7.943$\pm$0.040 mag \cr 
$w1$  & 7.651$\pm$0.024 mag \cr
$w2$ & 7.459$\pm$0.023 mag \cr
$w3$ & 7.252$\pm$0.019 mag \cr
$w4$ & 7.028$\pm$0.079 mag \cr
$M_{\rm J (2MASS)}$ & 10.77$^{+0.21}_{-0.25}$ mag \cr 
$M_{\rm H (2MASS)}$ & 10.11$^{+0.21}_{-0.25}$ mag \cr 
$M_{\rm Ks (2MASS)}$ & 9.75$^{+0.21}_{-0.25}$ mag \cr 
$M_{\rm bol}$ & 12.81$^{+0.23}_{-0.30}$ mag \cr 
Spectral Type &  M7V$\pm$0.5 \cr
T$_{\rm eff}$ & 2621$\pm$100\,K \cr
log\,(L/L$_{\odot})$ & --3.22$^{+0.09}_{-0.12}$ \cr 
Mass & 0.090$\pm$0.010 M$_{\odot}$ \cr

 \hline
 \label{tab_close_dwarf:properties}
 \end{tabular}
\end{table}
%

%
%
\section{Spectroscopic follow-up}
\label{close_dwarf:spectro_followup}
\subsection{Optical photometry and spectroscopy}
\label{close_dwarf:spectro_OPT}

We obtained imaging and spectroscopy of 2M1540 with the ESO Faint Object Spectrograph 
and Camera (EFOSC2) installed on the 3.58-m New Technology Telescope (NTT) in La Silla 
(Chile) during two consecutive nights in visitor mode on 4 and 5 October 2013\@. 
Both nights were dark, photometric with clear skies and sub-arcsec seeing.
EFOSC2 is equipped with a Loral/Lesser, thinned, ultraviolet flooded 2048$\times$2048
chip sensitive to optical wavelengths. The 2$\times$2 binning mode offers a pixel scale of 
0.24 arcsec and a field-of-view of 4.1$\times$4.1 arcmin.

We collected short exposures in the Sloan $i$ filter to improve our astrometric
measurements (right-hand side finding-chart in Fig.\ \ref{fig_close_dwarf:images}). 
We employed an on-source integration of 1.5 seconds in this filter.
On the other hand, we obtained low-resolution (R$\sim$500) optical spectroscopy with the 
grating number 16 set at parallactic angle to confirm the nature of our candidate. We used 
a 2\,sec exposure in the 
Gunn $i$ filter (\#705) centred on 669.4 nm to place our target in the slit and grism \#16 
with a slit of 1 arcsec for spectroscopic observations, covering the 600--1000 nm wavelength 
range. We obtained two single exposures of 300sec and 180s on 11 and 12 October 2013,
respectively, to check for spectral variability. We observed an internal flat field immediately after 
the spectrum of the target to correct for fringing, which affects grism \#16 beyond $\sim$720 nm.
We obtained bias and arc frames during the afternoon preceding our observations as well as
a spectro-photometric standard star \cite[LTT7379;]{hamuy92} during the night to apply the
instrumental response to our target.

\begin{figure*}
  \centering
  \includegraphics[width=0.49\linewidth, angle=0]{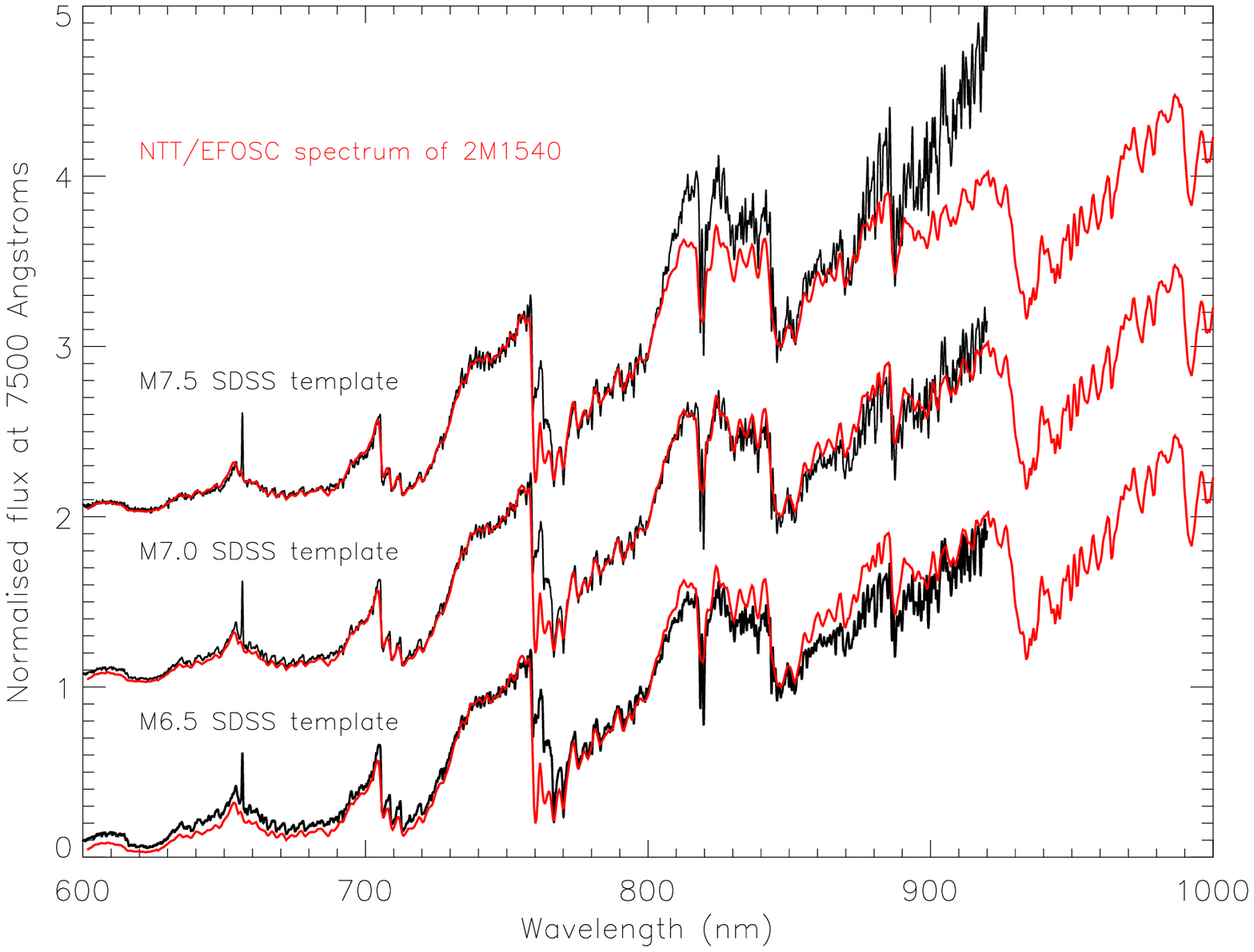}
  \includegraphics[width=0.49\linewidth, angle=0]{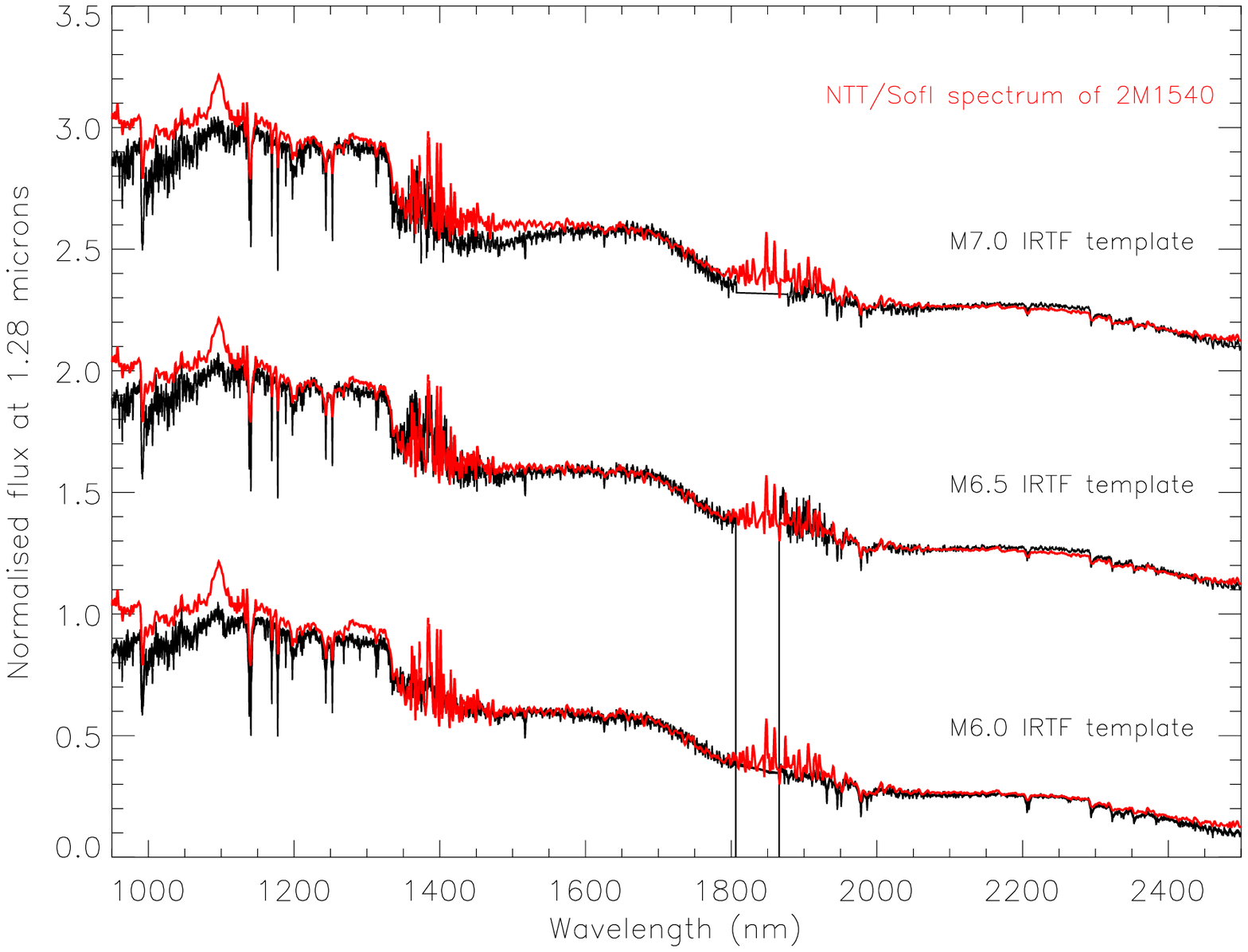}
   \caption{Low-resolution (R\,$\sim$\,500) optical (600--1000 nm) and
   near-infrared (900--2500 nm) spectra of 2M1540 (red lines) obtained with EFOSC2
   (left) and SofI (right) on the NTT, respectively. We normalised the optical
   and near-infrared spectra at 750 and 1650 nm, respectively.
   Plotted in black below the optical spectrum of 2M1540 are three spectral template 
   corrected for telluric absorptions (red line) from the SDSS spectroscopic
   database with spectral types bracketing our target \citep[M6.5, M7, and M7.5;][]{bochanski07a}. 
   Plotted in black below the near-infrared spectrum of 2M1540 are spectral templates 
   from the IRTF spectral library: Gl\,406 (M6), GJ\,1111 
   \citep[M6.5;][]{kirkpatrick91,jenkins09,kirkpatrick12}, and Gl\,644C
   \citep{zakhozhaj79,boeshaar85,kirkpatrick12}.
}
   \label{fig_close_dwarf:spectra}
\end{figure*}

We carried out the spectroscopic data reduction of 2M1540 under the IRAF 
environment \citep{tody86,tody93}. We used the imaging data only for astrometric
purposes. For the spectroscopy, we subtracted a median-combined bias frame and divided 
by a normalised internal flat field to remove fringing in the red part of the spectrum. 
We extracted a one-dimensional (1D) spectrum, using optimal aperture extraction and the {\it APALL} routine.
We calibrated our spectrum in wavelength with helium-argon arc lamps before applying the
instrumental response, using the LTT7379 spectro-photometric standard star observed each night.
The final 1D optical spectrum of 2M1540, normalised at 750 nm, is displayed on the 
left-hand side panel of Fig.\ \ref{fig_close_dwarf:spectra}.

Furthermore, we conducted photometric observations in the $V$, $R$, SDSS$r$, SDSS$i$, 
and SDSS$z$ filters with EFOSC2 on 13 March 2014\@.
We reduced the data using the ESO GASGANO pipeline on the mountain, which includes bias substraction 
and flat-field correction. Weather conditions were photometric and seeing varied between 
0.4 and 1.3 arcsec. We carried out aperture photometry with DAOPHOT within ithe IRAF environment. 
We converted the instrumental magnitudes into apparent magnitudes in the Cousin (Vega magnitudes) and Sloan (AB magnitudes) systems 
using photometric standards from \citet{landolt92} and \citet{smith02}, respectively
(see Table \ref{tab_close_dwarf:properties}).
The comparison of optical and near-infrared colours of 2M1540 with the colours of field M dwarfs 
published by \citet{kirkpatrick94} and \citet{west11} suggests an estimated spectral type of 
M6--M7, in agreement with our spectroscopic type determination of M7 (Section \ref{close_dwarf:spectro_SpT}).

%
%
%
\subsection{Near-infrared spectroscopy}
\label{close_dwarf:spectro_NIR}

We collected a low-resolution (R\,$\sim$\,600) near-infrared spectrum of 2M1540
with the blue (950--1640 nm) and red (1530--2520 nm) grisms combined with 
a slit of 1 arcsec with the Son of ISAAC (SofI) spectro-imager installed on the NTT
on 20 October 2013\@.
SofI is equipped with a Hawaii HgCdTe 1024$\times$1024 array with squared 18.5 micron 
pixels. We used the large-field mode, offering a field-of-view of 4.9$\times$4.9 arcmin 
with a 0.288 arcsec pixel scale.
We set the on-source integrations to five seconds, repeated five times with an ABBA pattern for 
both configurations to remove the sky contribution. We observed a hot star with the same
configuration\citep[HIP74940; $V$\,=\,8.58 mag; A0V;][]{houk78,vanLeeuwen07,hog00} 
immediately  before 2M1540 with an AB pattern in each configuration and at a similar 
airmass to correct for telluric features. 
The conditions were photometric at the time of the observations with sub-arcsec seeing. 
We observed Xenon arc lamps during the afternoon to calibrate our spectra in wavelength.

We used the ESO SofI pipeline (gasgano) on the mountain to obtain a combined dispersed image
from the four ABBA positions along the slit. We extracted a 1D spectrum with
standard spectroscopic routines under IRAF \citep{tody86,tody93} and calibrated it
in wavelengths with Xenon lamps with an rms better than 0.4\,\AA{} and 0.7\,\AA{}
in the blue and red part of the spectrum, respectively. Then, we divided this spectrum by the 
1D spectrum of the telluric standard (also reduced with the pipeline and extracted
with IRAF), which we later multiplied by the appropriate spectral template of an A0V star
smoothed to our resolution\footnote{www.eso.org/sci/observing/tools/standards/IR\_spectral\_library.html}.
The final near-infrared spectrum of 2M1540, normalised at 1650 nm, is displayed on
the right-hand side of Fig.\ \ref{fig_close_dwarf:spectra}.

%
%
%
\begin{figure*}
  \centering
  \includegraphics[width=\linewidth, angle=0]{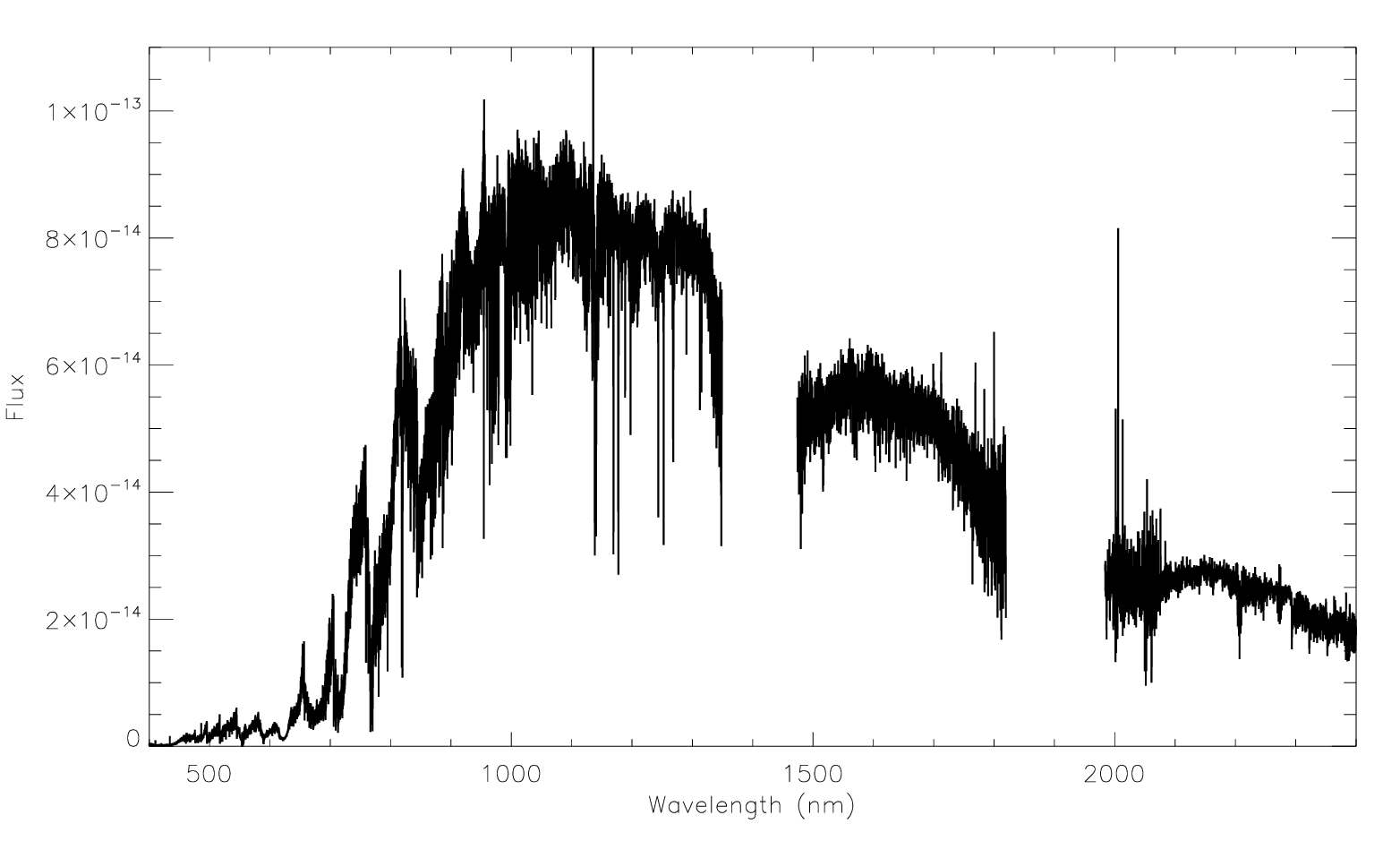}
   \caption{VLT/X-shooter spectrum of 2M1540 covering the UVB, VIS, and NIR
wavelength range, from 400 to 2500 nm at a resolution of 4000--6800\@. 
The spectrum has been corrected for telluric absorption.
}
   \label{fig_close_dwarf:XSHspec}
\end{figure*}
%

%
%
%
\begin{figure*}
  \centering
  \includegraphics[width=0.19\linewidth, angle=0]{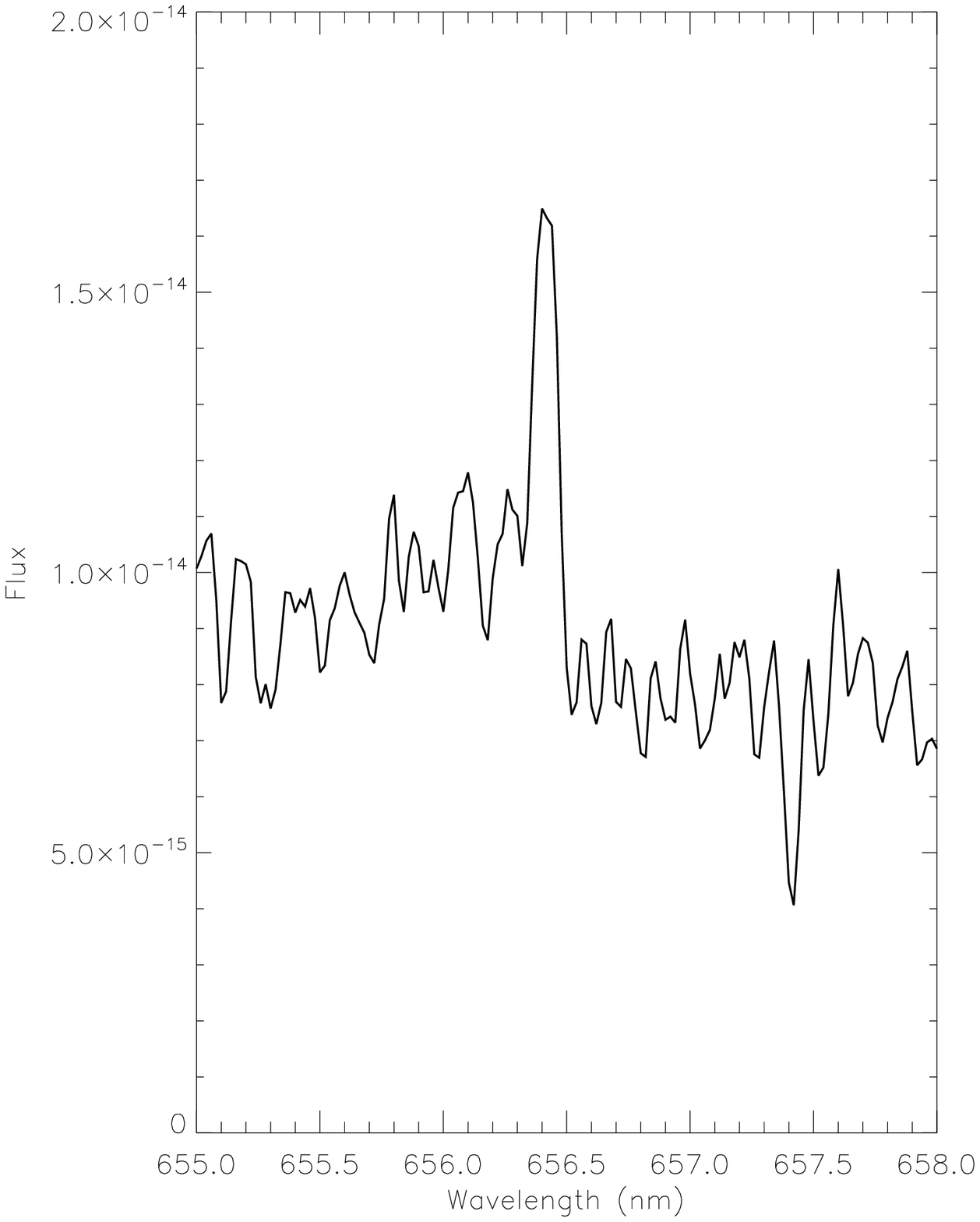}
  \includegraphics[width=0.19\linewidth, angle=0]{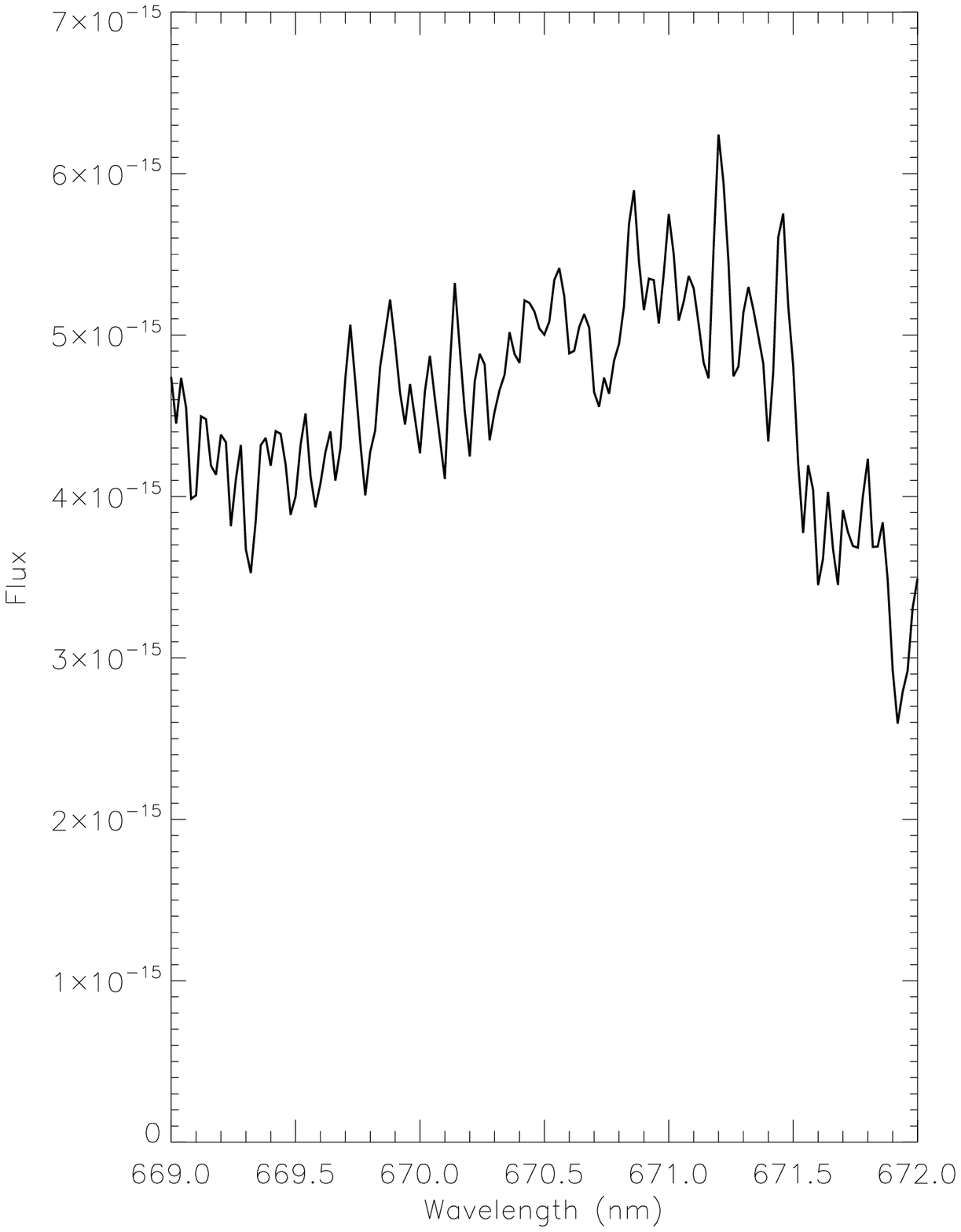}
  \includegraphics[width=0.19\linewidth, angle=0]{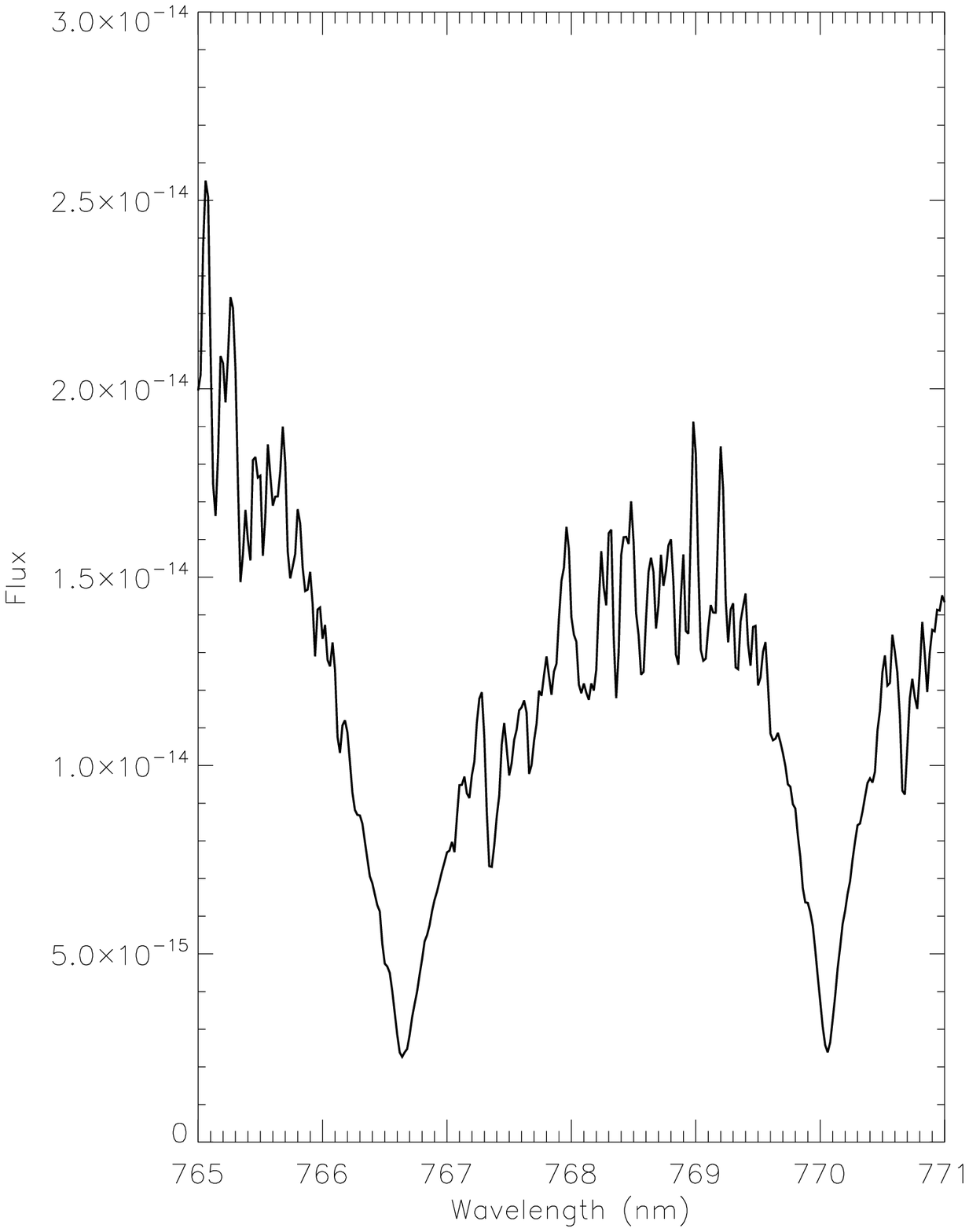}
  \includegraphics[width=0.19\linewidth, angle=0]{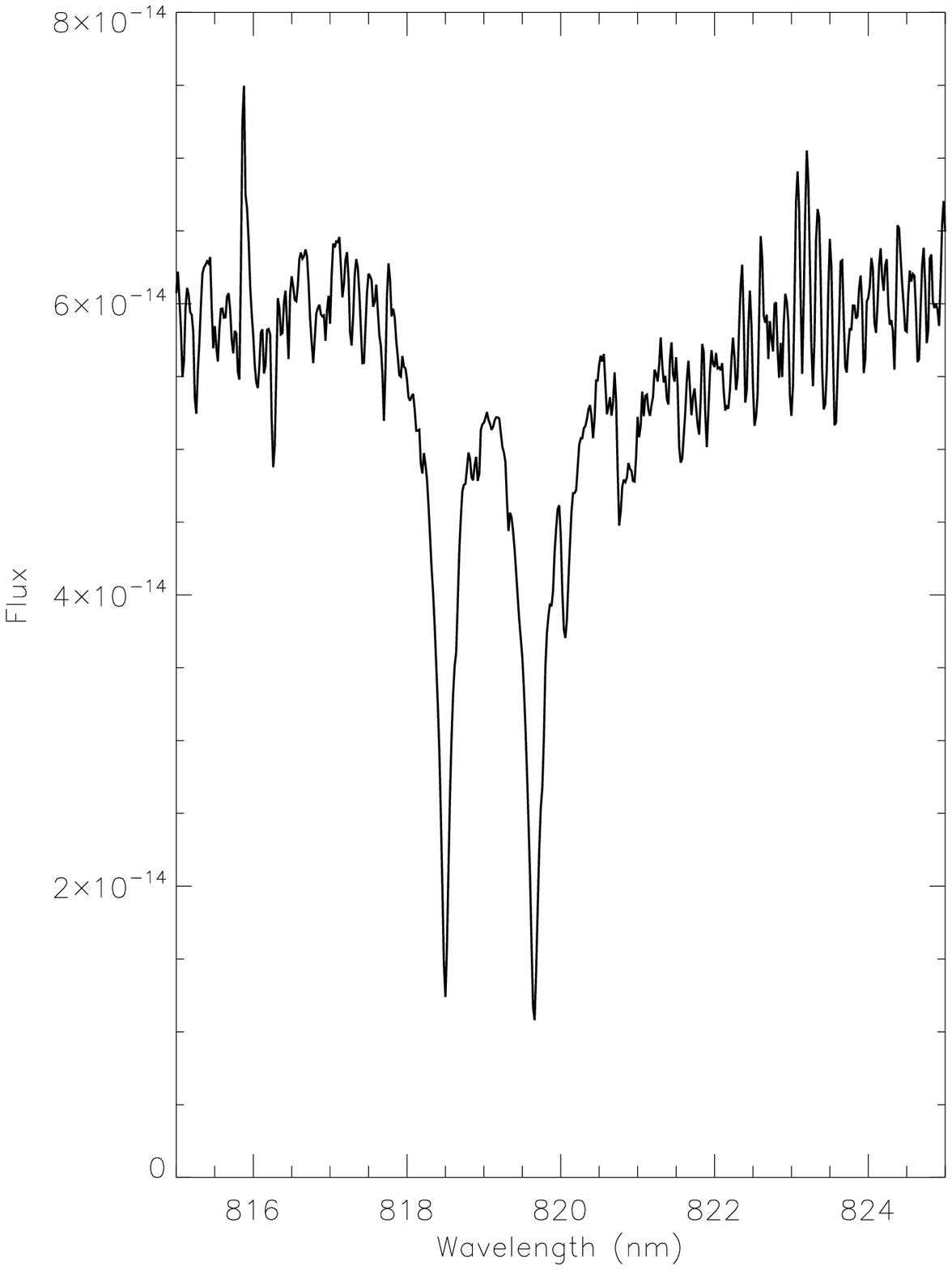}
  \includegraphics[width=0.19\linewidth, angle=0]{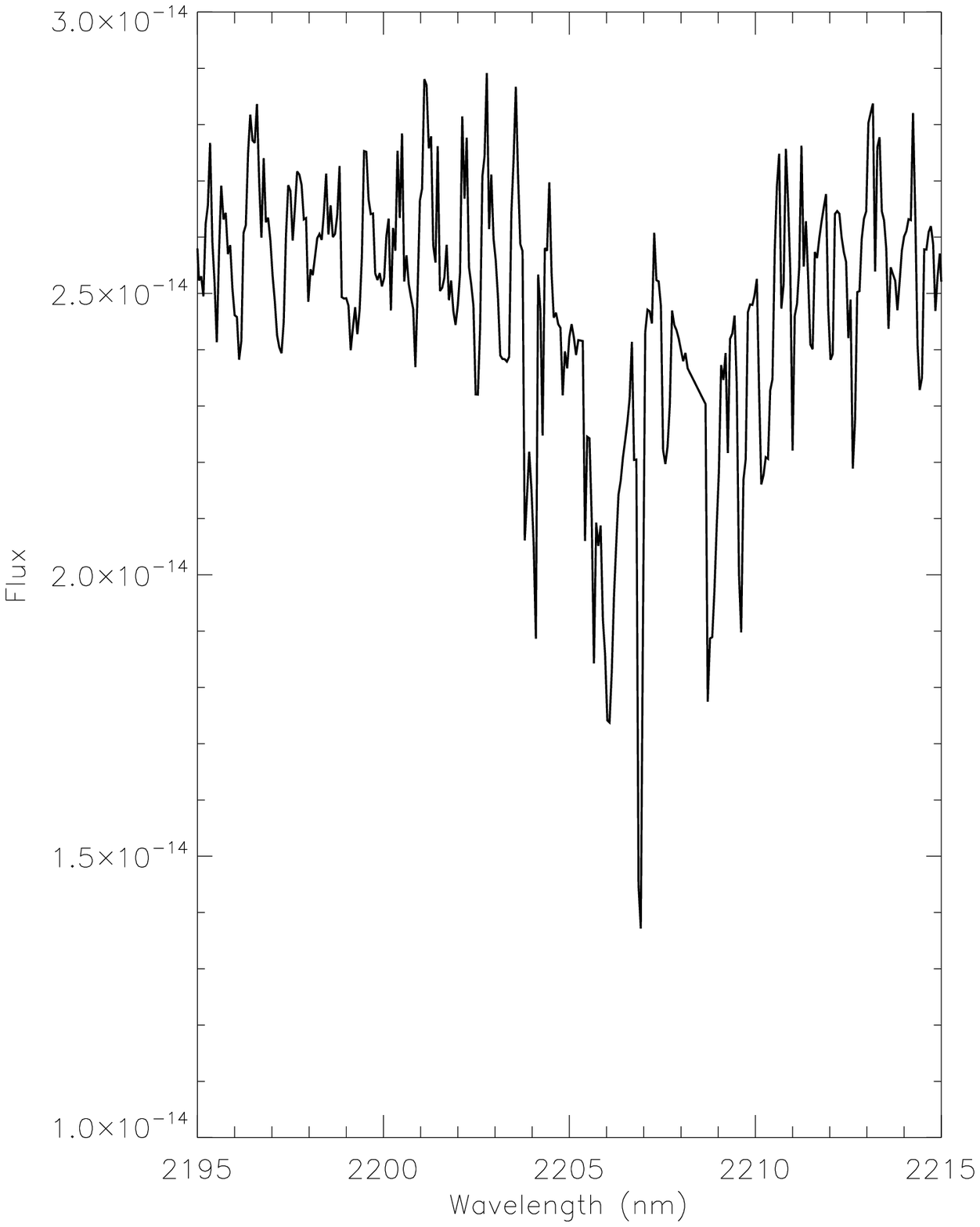}
   \caption{Zoom on some specific regions of the VLT/X-shooter spectrum 
of 2M1540: from left to right, H$\alpha$ at 656.3 nm, lithium at 670.8 nm,
potassium around 770 nm, and sodium at $\sim$819 nm and $\sim$2200 nm.
}
   \label{fig_close_dwarf:XSHspec_zoom}
\end{figure*}
%

%
%
\subsection{X-shooter spectrum}
We obtained a VLT/X-shooter spectrum of 2M1540 covering the ultraviolet, visible,
and near-infrared regions, from 350 to 2500 nm, at resolutions between 4000
and 7000 on 10 March 2014 in visitor mode (ESO programme number 092.C-0229). 
The sky was clear at the time of the observations with sub-arcsec seeing. 
We used four exposures of 120 sec in each arm, yielding a clear detection 
at wavelengths longwards of 400 nm due to the brightness of the object.
Calibrations were taken during the afternoon preceding the observations, in
accordance with the instrument calibration plan, including bias, dark, flat field,
and arc frames. Telluric \citep[Hip\,73881; B2--B3;][]{houk78,vanLeeuwen07}
and flux \citep[EG\,274; DA2;][]{oswalt88,hamuy92,holberg02,vanLeeuwen07} 
standards were observed very close in time to the target.

X-shooter \citep{vernet11} is a multi-wavelength cross--dispersed echelle spectrograph made
of three independent arms that simultaneously covers the ultraviolet (UVB;
300--560 nm), visible (VIS; 560--1020 nm), and near--infrared
(NIR; 1020--2480 nm) wavelength ranges thanks to two
dichroics that split the light. The spectrograph is equipped with three
detectors: a 4096$\times$2048 E2V CCD44-82, a 4096$\times$2048 MIT/LL
CCID\,20, and a 2096$\times$2096 Hawaii 2RG for the UVB, VIS, and NIR arms,
respectively. We used the 1.2 arcsec slit (1.3 arcsec for the UVB arm)
to achieve a nominal resolution of 4000 (8.1 pixel per full width at half maximum)
in the UVB, 6700 in the VIS arm (7.9 pixel per FWHM, and 3890 in the NIR arm
(5.8 pixel per full--width--half--maximum) in the VIS arm.

We reduced the X-shooter spectra on each arm independently using the command
lines part of the esorex package (version 2.2.0) provided by ESO\@.
We followed the steps enumerated in the manual to construct a final 2D spectrum
and the associated 1D spectrum for 2M1540 and the standard star in each arm.
First, we created a master bias and master dark. Afterwards, we determined
a first-guess order, and then we refined the line tables by illuminating the
X-shooter pinhole with a continuum lamp. Later, we created a master flat and
an order table tracing the flat edges before establishing the 2D map
of the instrument. Subsequently, we determined the efficiency of the whole
system made of the telescope, instrument, and detector. Finally, we
generated the 2D spectrum in nodding mode for 2M1540 and its standard star.
We also corrected the VIS and NIR parts of the X-shooter spectrum for telluric 
absorption using the normalised continuum of the telluric standard Hip\,73881\@.
In Fig.\ \ref{fig_close_dwarf:XSHspec}, we plot the 1D flux-calibrated
and telluric-corrected spectrum of 2M1540 covering the 350--2500 nm range. 
This spectrum is of high quality and will serve as template for future works 
because 2M1540 is currently the closest M7 dwarf to the Sun.
In Fig.\ \ref{fig_close_dwarf:XSHspec_zoom} we display some specific regions
around absorption/emission lines and doublets of interest.

%
%
\begin{table}
 \centering
 \caption[]{Pseudo-equivalent widths (pEWs) of lines and doublets measured in the
VLT/X-shooter spectrum of 2M1540\@.}
 \begin{tabular}{@{\hspace{0mm}}c @{\hspace{2mm}}c @{\hspace{2mm}}c @{\hspace{2mm}}c @{\hspace{2mm}}c @{\hspace{2mm}}l@{\hspace{0mm}}}
 \hline
 \hline
Line          &  Arm   &  Wavelength   & pEWs       \cr
 \hline
              &        &     nm        &  \AA{}     \cr
 \hline
Li            &  VIS   &   670.8        & $<$0.2  \cr
H$\alpha$     &  VIS   &   656.3        & $-$1.0$\pm$0.2 \cr
K{\small{I}}  &  VIS   &   766.5/769.9  & 7$\pm$1 / 4$\pm$1 \cr
Na{\small{I}} &  VIS   &   818.3/819.8  & 2.8$\pm$0.6 / 3.4$\pm$0.2 \cr
Na{\small{I}} &  NIR   &   1138/1140    & 2.8$\pm$0.2 / 3.4$\pm$0.3 \cr
K{\small{I}}  &  NIR   &   1169/1177    & 2.5$\pm$0.4 / 4.0$\pm$0.4 \cr
K{\small{I}}  &  NIR   &   1244/1253    & 1.7$\pm$0.3 / 2.2$\pm$0.4 \cr
\hline
\end{tabular}
 \label{tab_close_dwarf:XSH_EWs}
\end{table}

\subsection{Spectral classification}
\label{close_dwarf:spectro_SpT}

We attempted to classify 2M1540 using  the optical and infrared spectra independently.
The two optical spectra of 2M1540 taken on two consecutive nights are similar.
For the optical, we compared our EFOSC spectrum to the set of M dwarf templates
from the SDSS spectroscopic database \citep{york00}.
\citet{bochanski07a} made publicly available a repository of good-quality spectra
(corrected for telluric absorption) for M0--M9 dwarfs spanning the 380--940 nm wavelength
range at a resolution of 1800\footnote{The spectra used in this paper were downloaded
from www.astro.washington.edu/users/slh/templates/data/}.
The spectral types are derived from spectral indices defined in \citet{reid95}
and \citet{hawley02}. We created additional templates for the M0.5 to M8.5 subtypes
by averaging the optical spectra of the bracketing classes to fine-tune our
spectral classification. On the left-hand side panel of Fig.\ \ref{fig_close_dwarf:spectra}
we overplot our spectrum of 2M1540 (red line) on top of three Sloan spectral templates
with spectral types of M6.5, M7.0, and M7.5\@. The best fit is obtained for the M7 dwarf.
Hence, we optically classify our new discovery as an M7 dwarf, with an uncertainty of 0.5 subtype.
This classification is one subtype later than the spectral type reported by \citet{kirkpatrick14}.
We arrive at the same conclusion from the analysis of the X-shooter spectrum,
comparing with the spectrum of the nearby M6 dwarf Gl\,644C \citep{monet92,kirkpatrick12} 
kindly provided by the team that builds the X-shooter spectral library \citep{chen14a}.

Moreover, we computed spectral indices for the two spectra obtained on 4 and 5 October. 
We measured a PC3 index of 1.40--1.42, leading to spectral types of M6.2--M6.3$\pm$0.4 
\citep{martin99a}. We measured TiO5 and VO-a indices of 0.26--0.35 and 2.13--2.14 
\citep{reid95,kirkpatrick99}, yielding spectral types of M7.5--M8.2$\pm$0.4 and 
M6.1--M6.2$\pm$0.8, respectively. However, we emphasise that these indices are
not suitable for M6--M8 dwarfs due to the degeneracy around these spectral types,
as discussed by \citet{cruz02}. Overall, the average value of those three spectral
indices suggests a spectral type of M6.6$\pm$0.9, consistent with our classification
based on SDSS spectral templates.

We compared our SofI spectrum with several late-M dwarf templates from the 
NASA Infrared Telescope Facility (IRTF) Medium-Resolution Spectrograph and Imager 
\citep[SpeX;][]{rayner03,cushing04} 
library\footnote{http://irtfweb.ifa.hawaii.edu/$\sim$spex/IRTF\_Spectral\_Library/} 
\citep{cushing05,rayner09}. We plot three well-known M dwarfs (black spectra) below 
our SofI spectrum (red line) in Fig.\ \ref{fig_close_dwarf:spectra}: Gl\,406 (M6),
GJ\,1111 \citep[M6.5;][]{kirkpatrick91,jenkins09,kirkpatrick12}, and Gl\,644C 
\citep[M7;][]{zakhozhaj79,boeshaar85,kirkpatrick12}.
The M6.5 spectral template type seems to best reproduce the overall spectral energy 
distribution and the molecular absorption bands of 2M1540 in the 1100--2500 nm
range, in agreement with our optical classification. Based on the optical and 
near-infrared spectra, we finally adopt for our object a spectral type of M7.0$\pm$0.5
(Table \ref{tab_close_dwarf:properties}).

We did not detect lithium in absorption and set an upper limit of 0.2\AA{} on its
pseudo-equivalent width (Fig.\ \ref{fig_close_dwarf:XSHspec_zoom}).
We did not detect H$\alpha$ in emission nor lithium in absorption in the low-resolution
optical spectrum of 2M1540\@. However, we did see an emission line at 656.3 nm in
the X-shooter spectrum, with a pseudo-equivalent width of 1.0$\pm$0.2\AA{}.
(Fig.\ \ref{fig_close_dwarf:XSHspec_zoom}. The lack of detection in the Galaxy 
Evolution Explorer \citep[GALEX;][]{martin05a}, the Extreme Ultraviolet Explorer 
\citep[EUVE;][]{boyd94}, and the R\"ontgen satellite \citep[ROSAT;][]{truemper93}
catalogues corroborates the low level of activity in this M7 dwarf.

We measured the pseudo-equivalent widths of the potassium and sodium doublets
present in the optical and near-infrared regions of the X-shooter spectrum
with the {\tt{SPLOT}} routine in IRAF\@. We defined the pseudo-continuum 
immediately before and after the atomic absorption features at the spectral 
resolution of the data. We resolved the two doublets at 766.5/769.9 nm
818.3/819.8 nm, respectively. We measured pseudo-equivalent widths of
7$\pm$1\AA{}/4$\pm$1\,\AA{} and 2.8$\pm$0.6\,\AA{}/3.4$\pm$0.2\,\AA{}, respectively
(Table \ref{tab_close_dwarf:XSH_EWs}).
In the near-infrared, we measured pseudo-equivalent widths of 2.8$\pm$0.2/3.4$\pm$0.3\,\AA{},
2.5$\pm$0.4/4.0$\pm$0.4\,\AA{}, and 1.7$\pm$0.3/2.2$\pm$0.4\,\AA{} for the (resolved) 
sodium doublet at 1138/1140 nm and the potassium doublets at 1169/1177 and 1244/1253 nm, 
respectively (Table \ref{tab_close_dwarf:XSH_EWs}).
All these measurements are typical for old nearby late-M (M6.5--M8) dwarfs,
as measured on the IRTF spectral templates.

%
%
\begin{table}
 \centering
 \caption[]{Astrometry for 2M1540 from public catalogues and our own follow-up
 with the ESO NTT\@.}
 \begin{tabular}{@{\hspace{0mm}}c @{\hspace{2mm}}c @{\hspace{2mm}}c @{\hspace{2mm}}c @{\hspace{2mm}}c @{\hspace{2mm}}l@{\hspace{0mm}}}
 \hline
 \hline
 R.A.\             & dec                   & $\sigma_{R.A.}$ &  $\sigma_{dec}$ & MJD & Source \\
 \hline
 degrees    &     degrees   &  mas  &  mas  &    days       &               \\
 \hline
235.1661050 & $-$51.0250780 & 320  &  320  &  45078.690278 & GSC DSSir \\
235.1705326 & $-$51.0254346 & 350  &  350  &  46910.635417 & DSS-1  \\
235.1791324 & $-$51.0264342 & 112  &  103  &  50554.605556 & POSS2-Red \\
235.1809082 & $-$51.0265884 &  60  &   60  &  51364.024306 & 2MASS \\
235.1902161 & $-$51.0275764 &  71  &   67  &  55251.096000 & WISE   \\
235.1904811 & $-$51.0276193 &  44  &   41  &  55432.482000 & WISE3B \\
235.1932656 & $-$51.0278275 & 140  &  160  &  56570.003403 & NTT (2013) \\
235.1932680 & $-$51.0278335 & 120  &  110  &  56570.983900 & NTT (2013) \\
235.1937577 & $-$51.0279052 & 120  &  110  &  56728.342884 & NTT (2014) \\ 
\hline
\end{tabular}
 \label{tab_close_dwarf:parallax_data}
\end{table}

%
%
\begin{figure}[!h]
\centering
  \includegraphics[width=0.75\linewidth, angle=-90]{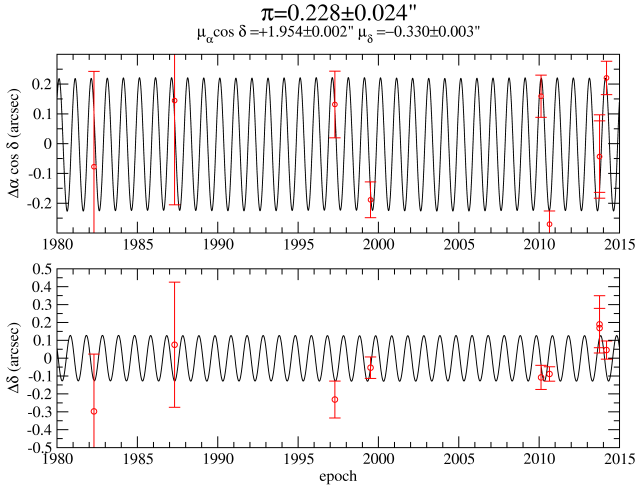}
   \caption{Parallactic motion for 2M1540\@. Proper motion produced by the fitting has been removed for clarity.
   }
   \label{fig_close_dwarf:parallax}
\end{figure}
%

%
%
\section{Distance estimates for 2M1540}
\label{close_dwarf:distance}
%

%
%
\subsection{Trigonometric parallax}
\label{close_dwarf:distance_PI}

We fitted for the parallax and proper motion of 2M1540 using the data in 
Table \ref{tab_close_dwarf:parallax_data}. For the DSS-1, POSS-2, and NTT images
(Fig.\ \ref{fig_close_dwarf:images}), 
we measured the centroids of all stars in each image using the task {\tt{daofind}} under
the IRAF environment \citep{tody86,tody93} and calculated the astrometric transformations to convert pixel 
coordinates to right ascension and declination
using the 2MASS Point Source Catalogue \citep{cutri03} as the frame of reference. For each frame, 
an initial astrometric transformation was fitted and then stars showing larger proper motions were 
discarded to obtain the final astrometric transformation. The number of astrometric reference stars 
varies betwen 50 and 100. Errors in coordinates were estimated after averaging the diferences between 
calculated positions of reference stars and their 2MASS coordinates.
We determined the parallax and proper motion using the method described 
in \citet{dupuy12}, although we did not perform a Markov chain Monte Carlo analysis.
We calculated the coordinates of Earth relative to the solar system barycentre with
the NOVAS routine \citep{kaplan12}.
We performed a least-squares fitting of both the proper motion and parallactic motion using the
MPFIT routine \citep{markwardt09}. We obtained a parallax of $\pi$\,=\,0.228$\pm$0.024 arcsec and 
a proper motion in each direction of $\mu_{\alpha}\cos{\delta}$\,=\,1.954$\pm$0.002 arcsec/yr 
and $\mu_\delta$\,=\,$-$0.330$\pm$0.003 arcsec/yr (Table \ref{tab_close_dwarf:properties};
Fig.\ref{fig_close_dwarf:parallax}). To check the consistency of this result we computed the parallax using 
several subsets of epochs from Table \ref{tab_close_dwarf:parallax_data}, and obtained similar estimations.
Our parallax measurement translates into a mean distance of 4.4$^{+0.5}_{-0.4}$\,pc.  
More data with new seeing-limited resolutions are still required for a finer adjustment 
of the distance until the GAIA satellite\footnote{More information at sci.esa.int/gaia/} 
provides us with exquisite parallax measurement.

We find that our parallactic distance, 4.4 pc, is closer than the distance ($\sim$6 pc)
reported by \citet{kirkpatrick14}. We re-computed the parallax with the MPFIT routine 
using the values given in Table 7 of \citet{kirkpatrick14} and derived a parallax of 
0.186$\pm$0.024 arcsec (d\,=\,5.38$^{+0.80}_{-0.61}$ pc), instead of 0.165$\pm$0.041 arcsec 
\citep[d\,=\,6.06$^{+2.0}_{-1.2}$ pc;][]{kirkpatrick14}. If we add our points from the 
NTT EFOSC2 epochs to the values of \citet{kirkpatrick14}, we find a parallax of 
0.208$\pm$0.020 arcsec (d\,=\,4.8$^{+0.5}_{-0.4}$ pc). The latter value is consistent with 
our previous determination. We find differences of up to 2 pc in the parallactic distances obtained by both groups,
suggesting that the error bars derived by the groups might be under-estimated if they have to be consistent with each other.
Despite this, the proper-motion measurements of Kirkpatrick et al. (2014) and ours are consistent whithin the 1-$\sigma$ the uncertainties quoted by the two groups.

%
%
\subsection{Spectrophotometric distance}
\label{close_dwarf:distance_SpT}

To estimate the spectroscopic distance of 2M1540, we selected the only two known
M7 dwarf within 8 pc, as listed in Table 4 of \citet{kirkpatrick12}: Gl\,644C (VB\,8) at
d\,=\,6.46$\pm$0.03 pc \citep{oppenheimer01,zakhozhaj79,boeshaar85} with
$J$\,=\,9.776$\pm$0.029 mag \citep{cutri03} and LHS\,3003 at d\,=\,6.88$\pm$0.16 pc 
\citep{reid02a,jenkins09b} with $J$\,=\,9.965$\pm$0.026 mag. Applying the standard 
transformation using those trigonometric parallaxes and infrared magnitudes for
M6.5 \citep{kirkpatrick12} and M7.5 dwarfs \citep{kirkpatrick94}, we infer spectroscopic 
distances of 4.43$^{+0.53}_{-0.37}$ and 4.33$^{+0.57}_{-0.41}$ pc using Gl\,644C 
and LHS\,3003, respectively. Taking the average of both estimates, we derive
a spectroscopic distance of 4.38 (3.92--4.96) pc for 2M1540 from the $J$-band, 
which agrees with our parallax estimate. Hence, we do not find evidence
for an equal-mass binary.

If, however, we consider a spectral type of M6.5, we derive a most probable range
of 4.43--5.61 pc for the spectroscopic distance, using GJ\,1111 and LHS\,292 as M6.5 templates 
at 3.62$\pm$0.04 and 4.54$\pm$0.07 pc, respectively \citep{vanLeeuwen07,kirkpatrick12}.
If we use the $K$-band and consider an error of half a subtype on the spectral
type, we derive a spectroscopic distance of 4.3$_{-0.2}^{+0.4}$ pc (1$\sigma$ error).
These distance estimates are within the error bars of our parallactic measurement.

We conclude that 2M1540 is the 
closest M7 dwarf to the Sun and the third of its subtype known within 8 pc.
Following Table 4 of \citet{kirkpatrick12}, 2M1540 would be among the 50
closest systems and would be between the 42$^{th}$ and 70$^{rd}$ place on the 
individual ranking taking into account the error bars on our parallax estimate. 
Our discovery and the recent anouncement of the closest
brown dwarf to the Sun \citep{luhman13a,mamajek13a,kniazev13,burgasser13b,boffin13} 
as well as the T7.5 dwarf WISE\,J0521$+$1025 at 5.0$\pm$1.3 pc \citep{bihain13a} 
demonstrate that the census of stars and brown dwarfs within 5 pc is still incomplete.

%
%
\section{Discussion}
\label{close_dwarf:discussion}

\subsection{Physical properties of 2M1540}
\label{close_dwarf:physics}

We have derived a spectral type for 2M1540 of M7 with an error of half a subtype
in both the optical and near-infrared spectra. This corresponds to an effective temperature 
of 2621$\pm$100\,K, adopting the temperature scale for field dwarfs 
from \citet{dahn02} and \citet{golimowski04a}. We have also inferred the luminosity of 
$\log L/L_{\odot}$\,=\,$-$3.22$^{+0.09}_{-0.12}$ for 2M1540 from its 2MASS $JHK_s$-band 
photometry, the bolometric correction from \citet{dahn02} and \cite{golimowski04a}, and 
our parallactic distance ($m-M$\,=$-$1.81$^{+0.25}_{-0.21}$). We can estimate the mass 
of 2M1540 by comparison of the derived luminosity with predictions from theoretical models 
\citep{baraffe03,burrows97} assuming an age of 1--10\,Gyr. We derive a mass of 
0.090$\pm$0.010 M$_{\odot}$, suggesting that 2M1540 is a very low-mass star, slightly 
above the hydrogen-burning mass limit.  

The lack of a strong feature due to lithium at 670.82 nm sets lower limits on the 
age and mass of the M7 dwarf. An upper limit of 0.1--0.2\,\AA{} on the lithium pEW implies 
a depletion factor larger than 1000 for an M7 source, according to the lithium curves 
of growth published in \citet{zapatero02c}. Additionally, 2M1540 is older than the 
Pleiades because lithium has been consumed from its photosphere 
\citep[M7 Pleiades dwarfs have lithium in their spectra, e.g.][]{rebolo96,martin98b,stauffer99}.
According to the lithium-depletion curves produced by \citet{baraffe98}, 2M1540 is most likely 
older than $\sim$600 Myr and has a mass higher than 0.06 M$_{\odot}$, in agreement 
with our previous derivation by comparison of the estimated dwarf effective temperature 
and luminosity with evolutionary models.

\subsection{Density of M7 dwarfs}
\label{close_dwarf:density}

From the 8 pc sample presented in \citet{kirkpatrick12}, we observe that only 
three objects have spectral types in the M7--M8 range (vB\,8, LHS\,3003, and Gl\,105\,C)
whose spectral types are estimated from broad band colours. Hence, the discovery of 
2M1540 represents an increase of 25\% in the number of these late M dwarfs known at 
distances closer than 8\,pc from our Sun. Using these four objects, we derive a density 
of $\rho$\,=\,1.9$\pm$0.9$\times$10$^{-3}$\,pc$^{-3}$ for M7 dwarfs in this volume, 
where error bars are given by Poisson count statistics. We should take this value as 
a lower limit, because other similar objects might have been missed in previous searches.
We also note that our discovery lies in the M7--M8 spectral range where the sample of \citet{cruz07} 
is highly incomplete because their colour cuts probably miss up to 40\% of 
these ultracool dwarfs. These authors estimated a number of 26.7 M7 dwarfs (after correction 
for incompleteness of objects with $J-K_{s}$ colours less than 1 mag) within 20\,pc in 36\% 
of the total sky area. These studies derived a density of 1.9--2.2$\times10^{-3}$\,pc$^{-3}$ 
for M7--M8 dwarfs, in agreement with our determination. Independent works by \citet{kirkpatrick94}
and \citet{caballero08e} derived similar densities of $\rho=1.91\times10^{-3}$\,pc$^{-3}$ for M7 dwarfs.

\section{Summary and future work}
\label{close_dwarf:conclusions}

Cross-matching the 2MASS and WISE public catalogues to identify high proper motion
low-mass stars and brown dwarfs, we independently discovered 2M1540 that we classified
spectroscopically as an M7$\pm$0.5 at $\sim$4.4 pc. We provided a high-quality
spectrum with moderate resolution. 2M1540 is the first M7 dwarf 
discovered within 5 pc and the third one within 8 pc. It is among the 50 nearest 
systems and among the 66 nearest stars.

This new M7 dwarf represents an ideal target for parallax follow-up, photometric 
variability studies, high-resolution spectroscopic follow-up to investigate the
role of clouds on the atmospheres of ultracool dwarfs, study the
membership to nearby moving groups, and search for
low-mass companions by direct imaging or radial velocity techniques.
This new discovery is the closest M7 dwarf in the sky, which makes it ideal to look
for substellar and planetary-mass companions including transiting 
Earth-like planets with a network of small telescopes
such as the Las Cumbres Observatory Global network \citep[LCOGT;][]{brown13a}
or radial velocity variations at optical and/or infrared wavelengths
with existing instruments such as ESO 3.6-/HARPS \citep{mayor03} and 
VLT/CRIRES \citep{kaeufl04}.

%
%
\begin{acknowledgements}
APG has been supported by Project No. 15345/PI/10 from the Fundaci\'on S\'eneca and the Spanish Ministry 
of Economy and Competitiveness (MINECO) under the grant AYA2010-21308-C03-03.
NL was funded by the Ram\'on y Cajal fellowship number 08-303-01-02\@.
This research has been supported by the Spanish Ministry of Economics and 
Competitiveness under the projects AYA2010-19136, AYA2010-21308-C3-02, 
AYA2010-21308-C03-03 and AYA2010-20535\@.
MTR acknowledge the support of the grant from CONICYT and the partial support 
from Center for Astrophysics FONDAP and Proyecto Basal PB06 (CATA).

This work is based on observations collected at the European Organisation 
for Astronomical Research in the Southern Hemisphere, Chile, under normal
programme numbers 092.C-0229(A) and 092.C-0229(B) in visitor mode.

This research has made use of the Simbad and Vizier databases, operated
at the Centre de Donn\'ees Astronomiques de Strasbourg (CDS), and
of NASA's Astrophysics Data System Bibliographic Services (ADS).

This research has made use of the USNOFS Image and Catalogue Archive
operated by the United States Naval Observatory, Flagstaff Station
 (http://www.nofs.navy.mil/data/fchpix/)
   
This publication makes use of data products from the Two Micron
All Sky Survey (2MASS), which is a joint project of the University
of Massachusetts and the Infrared Processing and Analysis
Center/California Institute of Technology, funded by the National
Aeronautics and Space Administration and the National Science Foundation.

This publication makes use of data products from the Wide-field Infrared
Survey Explorer, which is a joint project of the University of California,
Los Angeles, and the Jet Propulsion Laboratory/California Institute of
Technology, funded by the National Aeronautics and Space Administration.
\end{acknowledgements}
%

%
%
\bibliographystyle{aa}
\bibliography{../../AA/mnemonic,../../AA/biblio_old}
%

\end{document}